\documentclass[a4paper,twocolumn,english,aps,prl,superscriptaddress]{revtex4-1}
\usepackage[T1]{fontenc}
\usepackage[latin9]{inputenc}
\usepackage{color}
\usepackage{babel}
\usepackage{amsmath}
\usepackage{amssymb}
\usepackage{graphicx}
\usepackage{esint}
\usepackage[unicode=true,pdfusetitle,
 bookmarks=true,bookmarksnumbered=false,bookmarksopen=false,
 breaklinks=false,pdfborder={0 0 1},backref=false,colorlinks=false]
 {hyperref}
\usepackage{breakurl}

\makeatletter

\usepackage{babel}
\usepackage{babel}

\makeatother

\begin{document}

\title{Topological States in Two-Dimensional Su-Schrieffer-Heeger Models}

\author{Chang-An Li}
\email{changan.li@uni-wuerzburg.de}

\affiliation{Institute for Theoretical Physics and Astrophysics, University of
Würzburg, 97074 Würzburg, Germany}


\date{\today}
\begin{abstract}
We study the topological properties of the generalized two-dimensional
(2D) Su-Schrieffer-Heeger (SSH) models. We show that a pair of Dirac
points appear in the Brillouin zone (BZ), consisting a semimetallic
phase. Interestingly, the locations of these Dirac points are not
pinned to any high-symmetry points of the BZ but tunable by model
parameters. Moreover, the merging of two Dirac points undergoes a
novel topological phase transition, which leads to either a weak topological
insulator or a nodal-line metallic phase. We demonstrate these properties
by constructing two specific models, which we referred as type-I and
type-II 2D SSH models. The feasible experimental platforms to realize
our models are also discussed. 
\end{abstract}
\maketitle

\section{Introduction}

Topological phases of matter have attracted tremendous research interests
in recent decades \cite{QiXL11rmp,Kane10rmp}. Among those famous
topological models, the 1D Su-Schrieffer-Heeger (SSH) model provides
a prototype and simple model endowed with rich physics to investigate
topological phenomena in condensed matter physics \cite{SSH79prl}.
It exhibits fascinating topological properties such as the topological
phase transitions associated with Zak phase and fractional fermions
number at the ends of the sample \cite{SSH79prl}. It also helps to
clarify the theory of bulk polarization based on Berry phase \cite{King93prb},
which has wide and deep impacts on condensed matter physics in recent
decades, especially on the development of topological band insulators
\cite{SQS,BernevigBook}. 

Recently, the 1D SSH model has been extended to 2D on a square lattice.
For instance, Liu \textit{et al}. found that the 2D SSH model shows
nontrivial topological phases even the Berry curvature is zero in
the whole BZ \cite{LiuF17prl}. Benalcazar \textit{et al}. extended
the 1D SSH model to two-, and three-dimensional systems with a $\pi$-flux
inserted at each plaquette of the lattice. The proposed Benalcazar-Bernevig-Hughes
(BBH) models hold quantized bulk quadrupole and octupole moments in
2D and 3D, respectively \cite{Benalcazar17Science,BBH17prb}. Similar
to 1D SSH model, bound states carrying fractional charges exist at
the corners of the system. Thus the BBH provides a concert example
for the higher-order topological insulators (HOTIs). Such HOTIs generalize
the conventional bulk-boundary correspondence. Typically, a topological
bulk state in $d$-dimension has robust $(d-1)$-dimensional boundary
states. Nevertheless, HOTIs have localized states at boundaries that
are two or three dimensions lower than the bulk. The HOTIs have consequently
attracted both theoretical and experimental interest over past years
\cite{Langbehn17prl,SongZD17prl,Geier18prb,Schindler18NP,Serra-Garcia18nature,Peterson18nature,Ezawa18prl,Kudo19prl,Volpez19prl,XLSheng19prl,HuaCB20prb,ChenR20prl,ZhangRX20prl,Roy19prr,PengY19prl,LiC20prb,LiCA20prl,LiCA21prl,ChenXD19prl,QiY20prl,WeiQ21prl,NingZ22arxiv},
and the higher-order topological protection has been extended to superconductors
\cite{WangQ18prl,YanZB18prl,ZhangRX19prl,ZhangSB20prr,XYZhu18prb,WuYJ20prl,ZhangSB20prr2}
and semimetals \cite{WangHX20prl,Ghorashi20prl,WangZJ19prl}.

Since several types of 2D SSH models are possible when generalizing
the 1D SSH model, it is thus natural to ask whether these models exhibit
interesting topological properties. In this work, we investigate the
properties of two typical kinds of 2D SSH models. Remarkably, we find
that these models have rich topological phases. In the semimetallic
phase, a pair of Dirac points appear in the BZ. Interestingly, the
locations of the Dirac points are not pinned but can be easily tuned
by continuous parameter modulations without breaking any symmetries.
The merging of two Dirac points will experience a novel topological
phase transition which transform the system to either a weak topological
insulator or a nodal-line metallic phase. We demonstrate the topological
properties of these different phases by employing two independent
winding numbers together with boundary signatures and symmetry arguments.
We also discuss how to realize our model experimentally based on synthetic
quantum materials.

The remainder of this paper is organized as follows. Section II introduces
the type-I 2D SSH model and its band structure. Section III presents
the semimetallic phases of the type-I 2D SSH model. Section IV shows
the anisotropic nature of type-I 2D SSH model. Section V considers
properties of type-II 2D SSH model. Finally, we conclude our results
with a discussion in Section VI.

\section{type-I two-dimensional SSH model}

\begin{figure}
\includegraphics[width=1\linewidth]{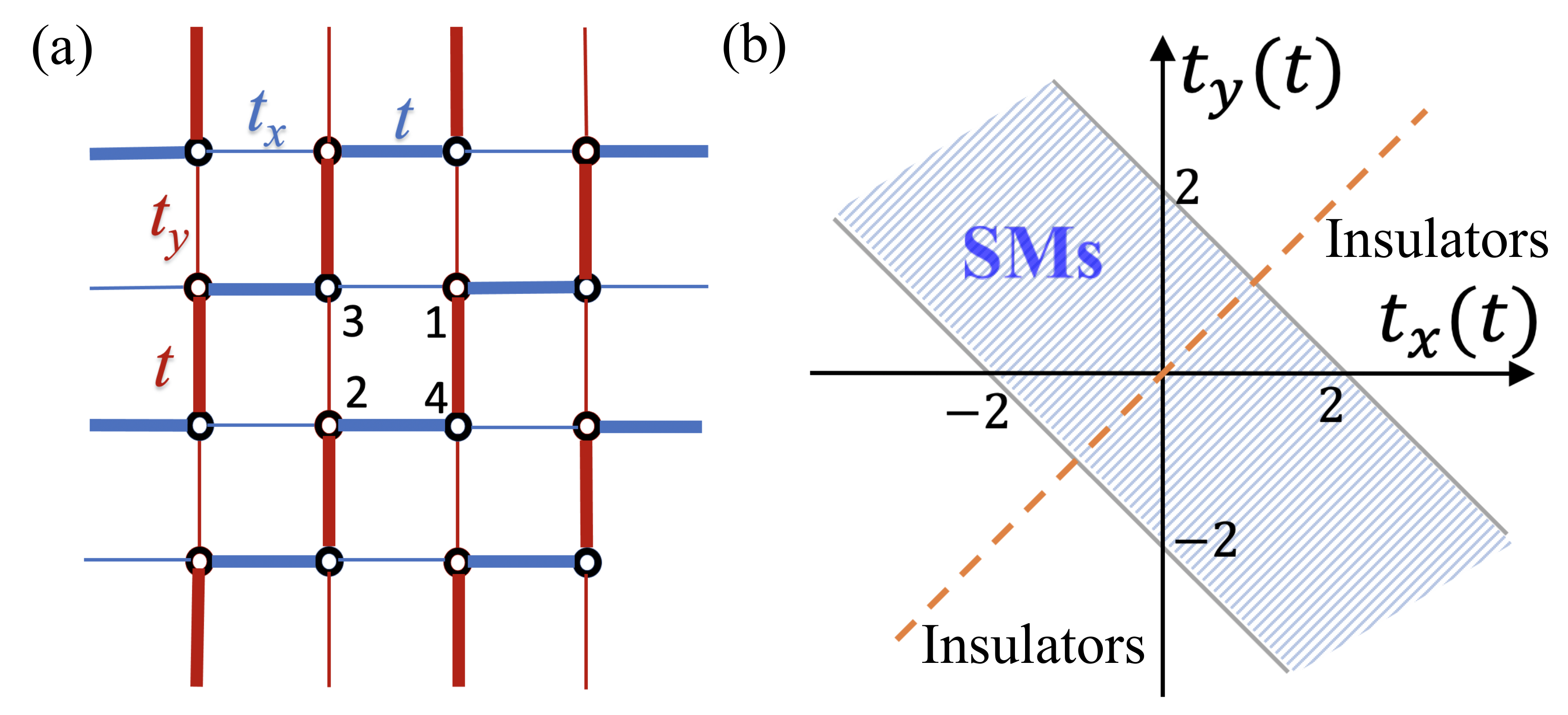}

\caption{(a) Schematic of the type-I 2D SSH lattice. Blue (red) thick and thin
bonds mark alternately dimerized hopping amplitudes in $x(y)$-direction.
(b) The full phase diagram of the type-I 2D SSH model in the parameter
space $(t_{x},t_{y})$. The shadowed region represents the semimetals
(SMs) with a pair of Dirac points. The orange dashed line at $t_{x}=t_{y}$
corresponds to a nodal-line metallic phase. Other regions are the
weak topological insulators. \label{fig:lattice-1}}
\end{figure}

Let us focus on the type-I 2D SSH model first \cite{LICA22arxiv}.
We consider a type-I 2D SSH model as shown in \textcolor{black}{Fig.\ \ref{fig:lattice-1}}(a),
where the weak (thin) bonds and strong (thick) bonds are alternately
dimerized along the two adjacent parallel lattice rows ($x$-direction)
or columns ($y$-direction). The four orbital degrees of freedom in
each unit cell are labeled as $1-4$. For clarity, we consider spinless
fermions. The lattice Hamiltonian is 
\begin{alignat}{1}
H_{1} & =\sum_{{\bf R}}(t_{x}C_{{\bf R},1}^{\dagger}C_{{\bf R},3}+tC_{{\bf R},2}^{\dagger}C_{{\bf R},4}+h.c.)\nonumber \\
 & +\sum_{{\bf R}}(tC_{{\bf R},1}^{\dagger}C_{{\bf R},4}+t_{y}C_{{\bf R},2}^{\dagger}C_{{\bf R},3}+h.c.)\nonumber \\
 & +\sum_{{\bf R}}(tC_{{\bf R},1}^{\dagger}C_{{\bf R}+\hat{x},3}+t_{x}C_{{\bf R},4}^{\dagger}C_{{\bf R}+\hat{x},2}+h.c.)\nonumber \\
 & +\sum_{{\bf R}}(t_{y}C_{{\bf R},1}^{\dagger}C_{{\bf R}+\hat{y},4}+tC_{{\bf R},3}^{\dagger}C_{{\bf R}+\hat{y},2}+h.c.),
\end{alignat}
where $C_{{\bf R},i}^{\dagger}$ is the creation operator for the
degree of freedom $i$ in the unit cell ${\bf R}$ with $i=1,2,3,4$,
as shown in the \textcolor{black}{Fig.\ \ref{fig:lattice-1}}(a).
Transforming it into the reciprocal space, the effective Bloch Hamiltonian
describing the type-I 2D SSH model reads

\begin{alignat}{1}
H_{1}({\bf k}) & =\left(\begin{array}{cc}
0 & q_{1}({\bf k})\\
q_{1}^{\dagger}({\bf k}) & 0
\end{array}\right),\label{eq:Hamiltonian}\\
q_{1}({\bf k}) & \equiv\left(\begin{array}{cc}
t_{x}+te^{ik_{x}} & t+t_{y}e^{ik_{y}}\\
t_{y}+te^{-ik_{y}} & t+t_{x}e^{-ik_{x}}
\end{array}\right);
\end{alignat}
where ${\bf k}=(k_{x},k_{y})$ is the 2D wave-vector; $t$ and $t_{x/y}$
are the staggered hopping amplitudes along $x/y$-directions. For
simplicity, we put the lattice constant to be unity and assume $t>0$
hereafter. From its off-diagonal form, the Hamiltonian in Eq.~\eqref{eq:Hamiltonian}
respects chiral (sublattice) symmetry. Explicitly, the chiral symmetry
is $\mathcal{C}H({\bf k})\mathcal{C}^{-1}=-H({\bf k})$ with the chiral-symmetry
operator $\mathcal{C}=\tau_{3}\otimes\sigma_{0}$, where $\tau$ and
$\sigma$ are Pauli matrices for different orbital degrees of freedom
in the unit cell. The energy bands and corresponding wave functions
can be obtained analytically. The energy bands of Eq.~\eqref{eq:Hamiltonian}
are 
\begin{equation}
E_{\eta}^{\pm}({\bf k})=\pm\sqrt{\xi_{\eta}^{2}({\bf k})+\zeta_{\eta}^{2}({\bf k})}=\pm|\varepsilon_{\eta}({\bf k})|,\label{eq:energy spectrum}
\end{equation}
where we have defined $\xi_{\eta}({\bf k})\equiv(t+t_{x})\cos\dfrac{k_{x}}{2}+\eta(t+t_{y})\cos\dfrac{k_{y}}{2}$,
$\zeta_{\eta}({\bf k})\equiv(t-t_{x})\sin\dfrac{k_{x}}{2}-\eta(t-t_{y})\sin\dfrac{k_{y}}{2}$,
and $\varepsilon_{\eta}({\bf k})\equiv\xi_{\eta}({\bf k})+i\zeta_{\eta}({\bf k})$
with $\eta=\pm1$. The convenient form of energy bands Eq.~\eqref{eq:energy spectrum}
will help us to locate the Dirac points and identify the phase diagram
of the system. 

\section{Semimetallic phases}

\begin{figure}
\includegraphics[width=1\linewidth]{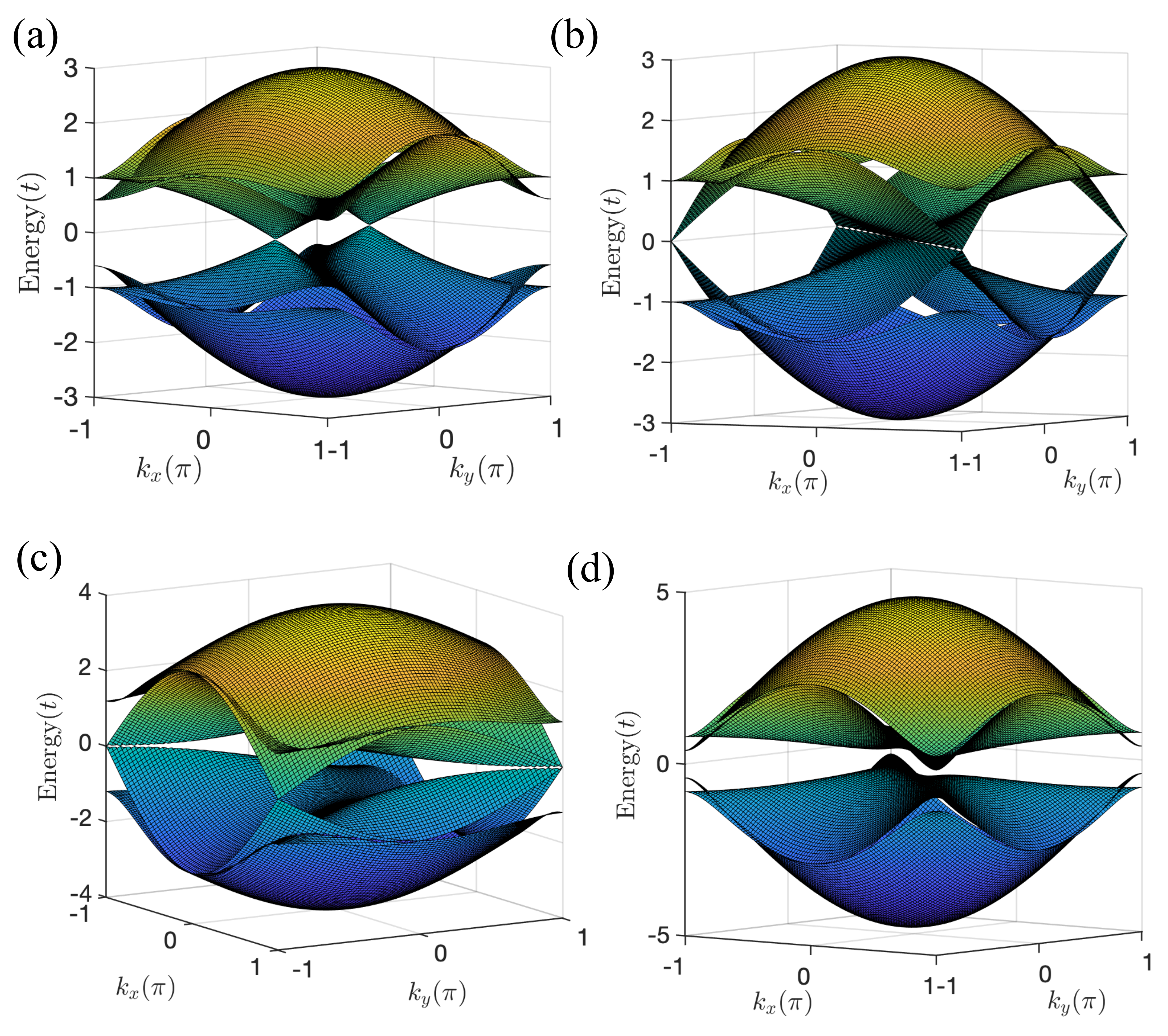}

\caption{Band structure in different phases of type-I 2D SSH model. (a) Band
structure in the semimetallic phase with $t_{x}=0.2t$ and $t_{y}=0.8t$.
(b) Band structure for the nodal-line metallic phase with $t_{x}=t_{y}=0.5t$.
(c) Band structure for a critical phase point with $t_{x}=1.6t$ and
$t_{y}=0.4t$. (d) Band structure for the weak topological insulators
with $t_{x}=1.6t$ and $t_{y}=1.2t$. \label{fig:threephases}}
\end{figure}

The type-I 2D SSH model actually possesses three different topological
phases, as shown in the phase diagram \textcolor{black}{Fig.\ \ref{fig:lattice-1}}(b).
Here we first discuss the semimetallic phase with a pair of Dirac
points within the region $|t_{x}+t_{y}|<2t$ and $t_{x}\neq t_{y}$.
Due to the presence of chiral symmetry, the conduction and valence
bands touch at zero energy [Fig. \ \ref{fig:threephases}(a)]. Thus, the existence of Dirac points is
constrained by the conditions $\xi_{\eta}({\bf k})=\zeta_{\eta}({\bf k})=0$.
Consequently, we find a pair of Dirac points located at ${\bf K}_{\pm}\equiv\pm(K_{x},-K_{y})$,
where $K_{x/y}$ are given by
\begin{equation}
K_{x/y}=2\arccos\sqrt{\frac{(t+t_{y/x})^{2}(2t-t_{x}-t_{y})}{4t(t^{2}-t_{x}t_{y})}}.\label{eq:Diracposition}
\end{equation}
Astonishingly, the Dirac points are not pinned to any high-symmetry
points but are highly tunable by parameter modulations. If we consider
a simple parameterization with $t_{x}=s\in[0,t]$, $t_{y}=t-s$, and
$t=1$, we find that the relation $K_{x}+K_{y}=2\pi/3$ holds true.
As a result, the Dirac points move along a line segment when we vary
the parameter $s$. Interestingly, no symmetries are broken as we
move around Dirac points by variation of $t_{x}$ and $t_{y}$. The
Dirac points are topologically protected by a quantized charge $Q_{{\bf K_{\pm}}}=\frac{1}{2\pi i}\oint_{\ell}d{\bf k}\cdot\mathrm{Tr}\left[q^{-1}({\bf k})\nabla_{{\bf k}}q({\bf k})\right]$,
where the loop $\ell$ is chosen such that it encircles a single Dirac
point ${\bf K}_{\pm}$ \cite{Schnyder11prb,Heikkila11jetp}. In essence,
it is based on the $\pi$ Berry phase, which is actually the same
as in graphene. The two Dirac points in the BZ have opposite topological
charges $Q_{{\bf K}_{\pm}}=\pm1$. They annihilate each other when
they meet in ${\bf k}$-space.

Let us then turn to the nodal-line metallic phase under the specific
condition $t_{x}=t_{y}$ {[}\textcolor{black}{Fig.\ \ref{fig:threephases}}(b){]}.
From Eq.~\eqref{eq:energy spectrum}, we find that the system exhibits
a gapless nodal line at
\begin{equation}
k_{x}+k_{y}=0,\ \mathrm{if}\ t_{x}=t_{y}\neq t.
\end{equation}
The appearance of a gapless nodal line is a direct consequence of
accidental mirror symmetry along the line $x+y=0$. In momentum space,
the mirror symmetry is expressed as $MH(k_{x},k_{y})M^{-1}=H(-k_{y},-k_{x})$.
Note that the Hamiltonian $H({\bf k})$ commutes with the mirror operator
$M$ along the nodal-line $k_{x}+k_{y}=0$. Therefore, we can label
the eigen states of the Hamiltonian $H({\bf k})$ by the eigen states
of mirror operator $M$ as 
\begin{alignat}{1}
H({\bf k})|\pm\rangle & =\pm E|\pm\rangle,M|\pm\rangle=\pm|\pm\rangle.
\end{alignat}
We further note that the mirror operator commute with the chiral symmetry
operator, i.e., $[\mathcal{C},M]=0$. Therefore, we can show that
$\mathcal{C}|+\rangle$ is also an eigenstate of $M$ with eigen value
$+1$. Moreover, $\mathcal{C}|+\rangle$ is eigenstate of $H({\bf k})$
with energy $+E$. Actually, the chiral symmetry maps the state $|+\rangle$
with energy $+E$ to state $\mathcal{C}|+\rangle$ with energy $-E$.
This implies that those states are degenerated states at energy $E=0$. 

\begin{figure}
\includegraphics[width=1\linewidth]{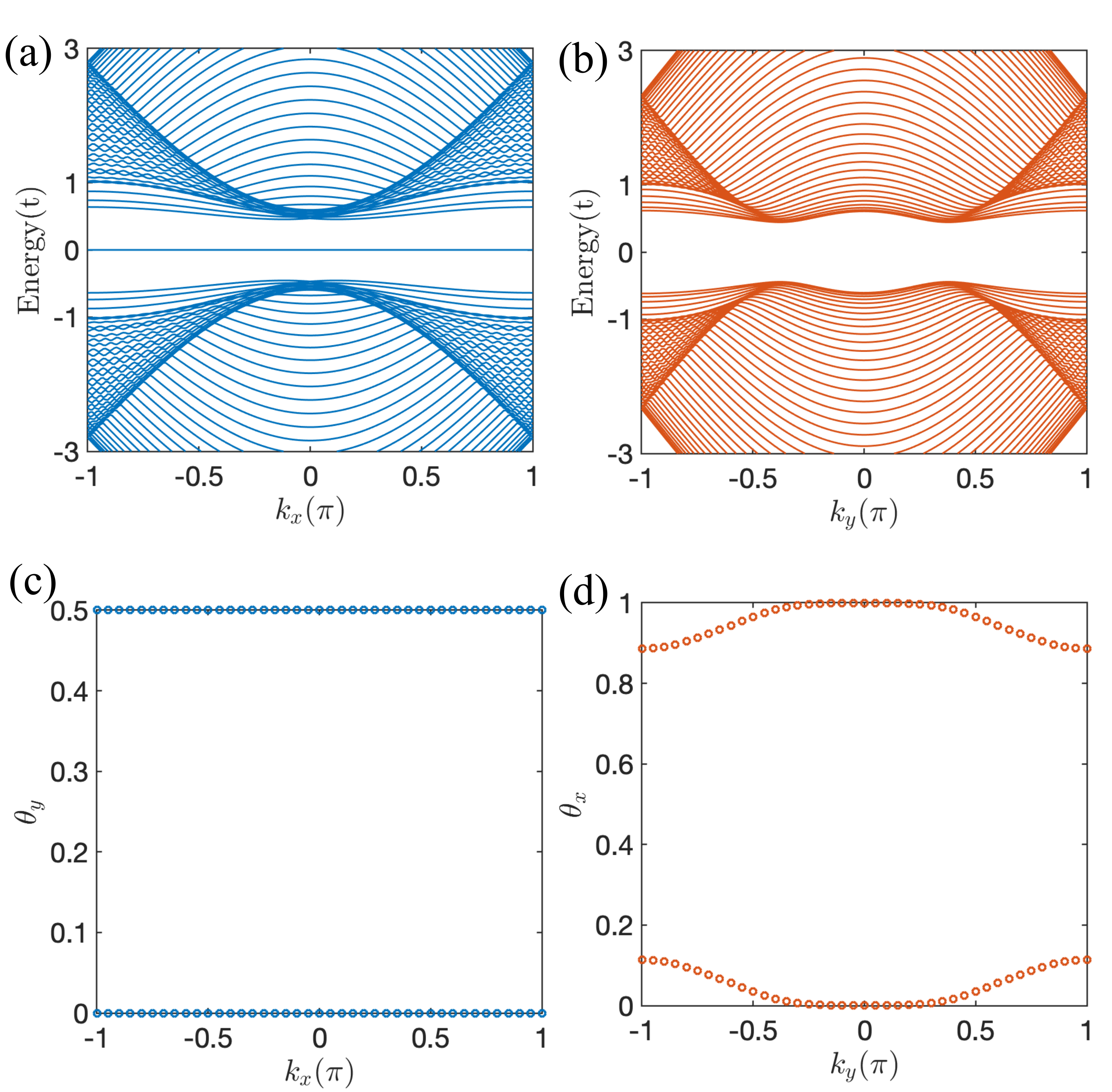}

\caption{(a) Energy spectrum of a ribbon along $x$-direction with width $W_{y}=20$.
Notice the flat band at zero energy. (b) Energy spectrum of the ribbons
along $y$-direction with width $W_{x}=20$. (c) Wannier bands $\theta_{y}$
as a function of $k_{x}$. (d) Wannier bands $\theta_{x}$ as a function
of $k_{y}$. The other parameters are $t_{x}=1.2t$ and $t_{y}=1.8t$.
\label{fig:anisotropic}}
\end{figure}

\section{Weak topological insulating phases}

The merging of two Dirac points can transfer the system from the semimetallic
phase to a weak topological insulator, which provides a novel type
of topological phase transition. Figure \textcolor{black}{\ref{fig:threephases}}(c)
presents the band structure at the critical merging points, at which
the spectrum stays linear along one direction while becomes parabolic
along another direction \cite{Montambaux09prb}. Specifically, the
weak topological insulators is located in the region $|t_{x}+t_{y}|>2t$
and $t_{x}\neq t_{y}$. The weak topological insulators possess a direct band gap, see Fig. \ \ref{fig:threephases}(d). It is described by two winding numbers $(w_{x},w_{y})$
with one of them being one and the other being zero. The winding number
is defined as
\begin{alignat*}{1}
w_{x/y} & =\frac{1}{2\pi i}\int_{0}^{2\pi}dk_{x/y}\mathrm{Tr}[q_{1}^{-1}({\bf k})\partial_{k_{x/y}}q_{1}({\bf k})]
\end{alignat*}
for arbitrary $k_{y/x}\in[0,2\pi]$. Actually, this weak topological
insulators can be further divided into two subphases: (i) $w_{x}=1,w_{y}=0$
($t_{x}>t_{y}$ and $|t_{x}+t_{y}|>2t$) and (ii) $w_{x}=0,w_{y}=1$
($t_{x}<t_{y}$ and $|t_{x}+t_{y}|>2t$). When $w_{x}=1,w_{y}=0$
($w_{x}=0,w_{y}=1$), the system is nontrivial along $x$($y$)-direction
and trivial along $y(x)$-direction. It is clear that once crossing
the boundary $t_{x}=t_{y}$ the system will shift from subphase (i)
to subphase (ii) or vice versa. Correspondingly, a totally flat edge
band exists in the gap of the energy spectrum of a ribbon along $x$($y$)-direction
for the subphase (i) (subphase (ii)). Figures \textcolor{black}{\ref{fig:anisotropic}}(a)
and (b) present the band structure of a ribbons along $x$- and $y$
-direction, respectively, for the subphase (ii). The flat edge bands
exist only in Fig. \textcolor{black}{\ref{fig:anisotropic}}(a). Notably,
neither the topologically trivial insulator with $w_{x}=w_{y}=0$
nor the nontrivial phase with $w_{x}=w_{y}=1$ appear in the inclined
2D SSH model.

Furthermore, the calculation of Wannier bands can also provide consistent
results with that of $w_{x/y}$ to identify the topological properties.
Specifically, the Wilson loop operator parallel to $y$ direction
is constructed as \cite{BBH17prb,Alexandradinata14prb}

\begin{equation}
\hat{P}_{y,{\bf k}}=P_{N_{y}\delta k_{y}+k_{y}}P_{(N_{y}-1)\delta k_{y}+k_{y}}\cdots P_{\delta k_{y}+k_{y}}P_{k_{y}},\label{eq:Wilsonloop}
\end{equation}
where each projection operator is defined as $P_{m\delta k_{y}+k_{y}}\equiv\sum_{n\in N_{\mathrm{occ}}}|u_{k_{x},m\delta k_{y}+k_{y}}^{n}\rangle\langle u_{k_{x},m\delta k_{y}+k_{y}}^{n}|$
with $|u_{k_{x},m\delta k_{y}+k_{y}}^{n}\rangle$ being the $n$-th
eigen state of occupied bands at point $(k_{x},m\delta k_{y}+k_{y})$,
and $m$ is an integer taking values from $\{1,2,\cdots,N_{y}\}$.
The projection method can avoid the arbitrary phase problem in numerical
realizations. Here $N_{y}$ is the number of unit cells, $n$ is the
band index, and $N_{\mathrm{occ}}$ is the number of occupied bands.
Note that $\hat{P}_{y,{\bf k}}$ has dimension of $N$ now with $N$
being the total bands number. After projection onto the occupied bands
at base point ${\bf k}$, there is $N_{\mathrm{occ}}\times N_{\mathrm{occ}}$
matrix $\mathcal{W}_{y,{\bf k}}$ that defines a Wannier Hamiltonian
$H_{\mathcal{W}_{y}}({\bf k})$ from the relation $\mathcal{W}_{y,{\bf k}}=\exp[iH_{\mathcal{W}_{y}}({\bf k})]$.
The eigen values of $H_{\mathcal{W}_{y}}({\bf k})$ give the Wannier
bands $2\pi\theta_{y}(k_{x})$ associated with eigen states $|\theta_{y,{\bf k}}^{j}\rangle$,
$j\in\{1,2,\cdots,N_{\mathrm{occ}}\}$. The Wannier bands plotted
in Figs. \textcolor{black}{\ref{fig:anisotropic}}(c) and (d) are
corresponding to the cases in Figs. \textcolor{black}{\ref{fig:anisotropic}}(a)
and (b). It is clear the two occupied bands in Fig. \textcolor{black}{\ref{fig:anisotropic}}(c)
gives a quantized half-integer polarization while the two occupied
bands in Fig. \textcolor{black}{\ref{fig:anisotropic}}(d) gives a
zero polarization (mod 1). The quantized half-integer polarization
indicates the nontrivial topological properties. 

\section{type-II two-dimensional SSH model}

\begin{figure}
\includegraphics[width=1\linewidth]{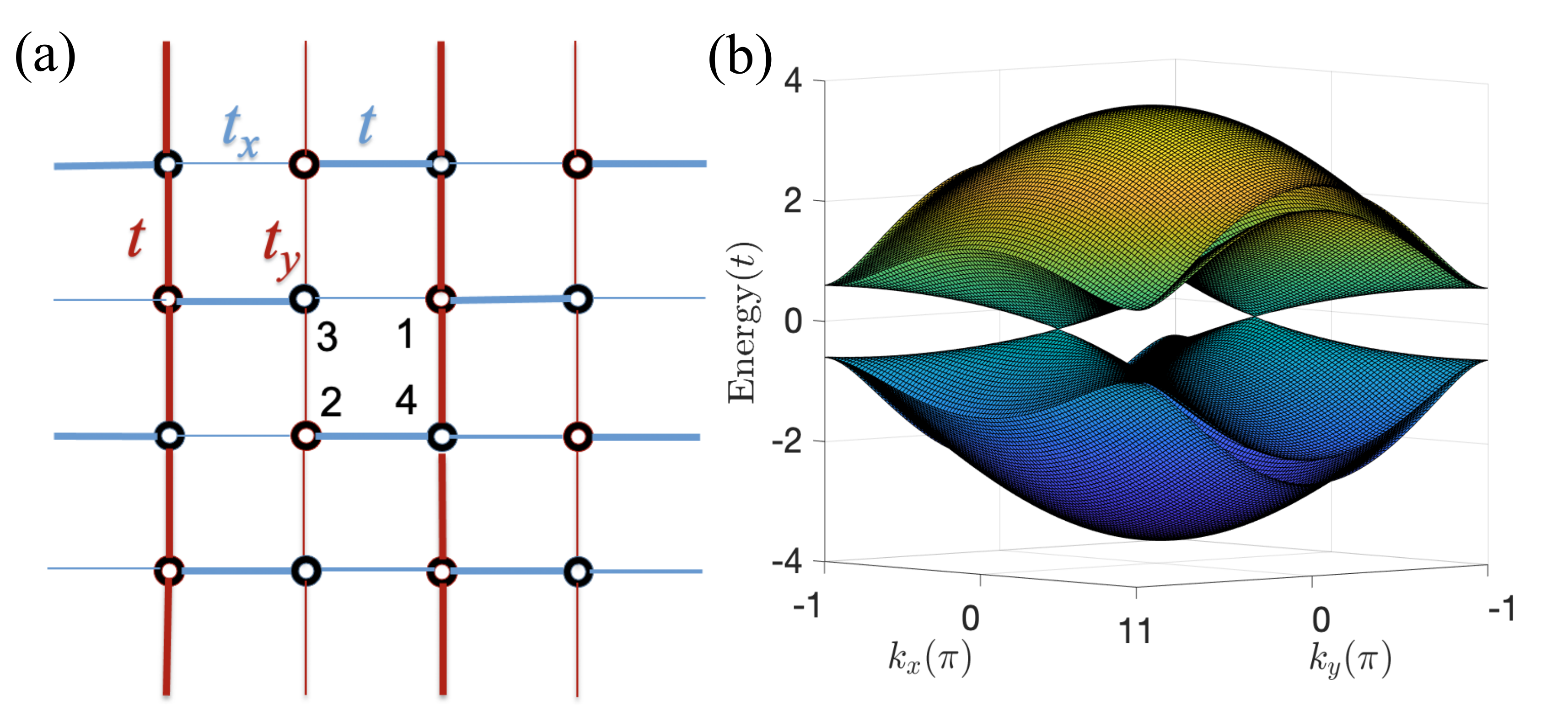}

\caption{(a) Schematic of the type-II 2D SSH lattice. Blue (red) thick and
thin bonds mark alternately dimerized hopping amplitudes in $x(y)$-direction.
(b) The band structure with at pair of Dirac points. Her we take $t_{x}=0.4t$,
$t_{y}=1.2t$. \label{fig:lattice-2}}
\end{figure}
Now, let us consider another similar model: the type-II 2D SSH model,
in which the alternatively dimerization pattern is shown in \textcolor{black}{Fig.\ \ref{fig:lattice-2}(a)}.
The lattice Hamiltonian reads as 

\begin{alignat}{1}
H_{2} & =\sum_{{\bf R}}(t_{x}C_{{\bf R},1}^{\dagger}C_{{\bf R},3}+tC_{{\bf R},2}^{\dagger}C_{{\bf R},4}+h.c.)\nonumber \\
 & +\sum_{{\bf R}}(tC_{{\bf R},1}^{\dagger}C_{{\bf R},4}+t_{y}C_{{\bf R},2}^{\dagger}C_{{\bf R},3}+h.c.)\nonumber \\
 & +\sum_{{\bf R}}(tC_{{\bf R},1}^{\dagger}C_{{\bf R}+\hat{x},3}+t_{x}C_{{\bf R},4}^{\dagger}C_{{\bf R}+\hat{x},2}+h.c.)\nonumber \\
 & +\sum_{{\bf R}}(tC_{{\bf R},1}^{\dagger}C_{{\bf R}+\hat{y},4}+t_{y}C_{{\bf R},3}^{\dagger}C_{{\bf R}+\hat{y},2}+h.c.).
\end{alignat}
The type-II model has many similarities with the type-I model, thus
we just focus on the semimetallic phase with Dirac points here. The
effective Bloch Hamiltonian describing the type-II 2D SSH model has
the same form as Eq.~\eqref{eq:Hamiltonian} but with the off-diagonal
parts replaced as 

\begin{alignat}{1}
q_{2}({\bf k}) & \equiv\left(\begin{array}{cc}
t_{x}+te^{ik_{x}} & t+te^{ik_{y}}\\
t_{y}+t_{y}e^{-ik_{y}} & t_{x}+te^{-ik_{x}}
\end{array}\right).
\end{alignat}
Its energy bands are 
\begin{equation}
E_{\eta}^{\pm}({\bf k})=\pm\sqrt{h_{0}({\bf k})+\eta\sqrt{\sum_{j=x,y,z}h_{j}^{2}({\bf k})}},\label{eq:energy spectrum-2}
\end{equation}
where we have defined the functions as $h_{0}({\bf k})\equiv(t-t_{x})^{2}+4t_{x}t\cos^{2}\frac{k_{x}}{2}+2(t^{2}+t_{y}^{2})\cos^{2}\dfrac{k_{y}}{2}$,
$h_{x}({\bf k})\equiv2\cos\frac{k_{y}}{2}\left[t(t_{x}+t_{y})\cos(k_{x}+k_{y}/2)+(t^{2}+t_{x}t_{y})\cos\frac{k_{y}}{2}\right]$,
$h_{y}({\bf k})\equiv2\cos\frac{k_{y}}{2}\left[t(t_{x}+t_{y})\sin(k_{x}+\frac{k_{y}}{2})+(t^{2}+t_{x}t_{y})\sin\frac{k_{y}}{2}\right]$,
and $h_{z}({\bf k})\equiv2(t^{2}-t_{y}^{2})\cos^{2}\dfrac{k_{y}}{2}$.
The type-II model has a glide-mirror symmetry: performing a mirror
symmetry $M_{x}$ and then a half translation $g_{y}$ along $y$-direction,
the system goes back to itself.

Its Dirac points are located along $k_{x}=0$ (or $k_{x}=\pi$) when
$t_{y}>0$ (or $t_{y}<0$) [see Fig. \ \ref{fig:lattice-2}(b)]. Explicitly, the Dirac points locate at
$(0,\pm2\arccos\sqrt{\frac{(t+t_{x})^{2}}{4tt_{y}}})$ for $t_{y}>0$
or $(\pi,\pm2\arccos\sqrt{\frac{(t-t_{x})^{2}}{-4tt_{y}}})$ for $t_{y}<0$.
Corresponding, the physical solutions hold under the condition $(t+t_{x})^{2}<4t_{y}t$
or $(t-t_{x})^{2}<-4t_{y}t$. The effective Hamiltonian close to the
Dirac points can also be obtained analytically. For simplicity, let
us focus on the case of $t_{y}>0$. To this end, we need to get the
two zero-energy eigen states at the Dirac points as a basis and then
project the full Hamiltonian to the basis. Finally, the effective
Hamiltonian is expressed as 
\begin{equation}
H_{\mathrm{eff}}({\bf k})=v_{x}\kappa_{x}\sigma_{x}-v_{y}\kappa_{y}\sigma_{y},
\end{equation}
where $v_{x}=\frac{\sqrt{t_{y}t}(t-t_{x})}{t+t_{y}}$, and $v_{y}=\mathrm{sgn}(t+t_{x})\frac{\sqrt{t_{y}t(4t_{y}t-(t+t_{x})^{2})}}{t+t_{y}}.$

\section{Discussion and conclusions }

Here we discuss how to realize our proposals experimentally. The most
important ingredient is the controllable nearest-neighbor couplings
between sites on the square lattice. Fortunately, such techniques
have been developed in synthetic quantum materials such as photonic
and acoustic crystals \cite{WangZ09nature,XieBY19prl,ChenXD19prl,Serra-Garcia18nature,Ni19nm},
electric circuits \cite{Imhof18np}, and waveguides \cite{Peterson18nature,Cerjan21prl}.
For instance, to realize our model in an acoustic system, the 3D printed
``atoms'' can be arranged to a square lattice with four contained
in each unit cell and the alternately dimerized couplings between
neighbours can be modulated the diameters that the sound wave go through.
Another feasible platform to realize our model is based on ultracold
gases in optical lattices \cite{Christian17Science,Tarruell12nature},
in which the lattice geometry and hopping strengths are adjustable.

Note that our results are distinctively different from recent reports
to realize Dirac states in square lattices \cite{Shao21prl,Xue21arXiv}.
These proposals require necessary $\pi$ fluxes on each plaquette,
and the Dirac points are pinned to boundaries of the BZ, which may
makes it more difficult to detect experimentally. While our 2D SSH
model does not require delicate manipulations of external flux. Interestingly,
our models even provide platforms to realize the so called toric-code
insulator \cite{Tam22prb}.

In conclusion, we have proposed the 2D SSH models on a square lattice
to realize tunable Dirac states. We have found that the locations
of Dirac points are not pinned in the BZ but movable by parameter
modifications. The merging of two Dirac points leads to a topological
phase transition, which converts the system from a semimetallic phase
to either a nodal-line metallic or a weak topological insulator. We
expect that our model can be realized in different metamaterial platforms.

\section{Acknowledgments}

The author acknowledges S. B. Zhang, S. J. Choi, B. Fu, and B. Trauzettel for helpful discussions. This work was supported by the DFG (SPP1666
and SFB1170 ``ToCoTronics''), the Würzburg-Dresden Cluster of Excellence
ct.qmat, EXC2147, Project-id 390858490, and the Elitenetzwerk Bayern
Graduate School on ``Topological Insulators'', and the NSF of Zhejiang
under Grant No. LQ20A04005.


\begin{thebibliography}{56}%
\bibliographystyle{apsrev4-1}
\makeatletter
\providecommand \@ifxundefined [1]{%
 \@ifx{#1\undefined}
}%
\providecommand \@ifnum [1]{%
 \ifnum #1\expandafter \@firstoftwo
 \else \expandafter \@secondoftwo
 \fi
}%
\providecommand \@ifx [1]{%
 \ifx #1\expandafter \@firstoftwo
 \else \expandafter \@secondoftwo
 \fi
}%
\providecommand \natexlab [1]{#1}%
\providecommand \enquote  [1]{``#1''}%
\providecommand \bibnamefont  [1]{#1}%
\providecommand \bibfnamefont [1]{#1}%
\providecommand \citenamefont [1]{#1}%
\providecommand \href@noop [0]{\@secondoftwo}%
\providecommand \href [0]{\begingroup \@sanitize@url \@href}%
\providecommand \@href[1]{\@@startlink{#1}\@@href}%
\providecommand \@@href[1]{\endgroup#1\@@endlink}%
\providecommand \@sanitize@url [0]{\catcode `\\12\catcode `\$12\catcode
  `\&12\catcode `\#12\catcode `\^12\catcode `\_12\catcode `\%12\relax}%
\providecommand \@@startlink[1]{}%
\providecommand \@@endlink[0]{}%
\providecommand \url  [0]{\begingroup\@sanitize@url \@url }%
\providecommand \@url [1]{\endgroup\@href {#1}{\urlprefix }}%
\providecommand \urlprefix  [0]{URL }%
\providecommand \Eprint [0]{\href }%
\providecommand \doibase [0]{http://dx.doi.org/}%
\providecommand \selectlanguage [0]{\@gobble}%
\providecommand \bibinfo  [0]{\@secondoftwo}%
\providecommand \bibfield  [0]{\@secondoftwo}%
\providecommand \translation [1]{[#1]}%
\providecommand \BibitemOpen [0]{}%
\providecommand \bibitemStop [0]{}%
\providecommand \bibitemNoStop [0]{.\EOS\space}%
\providecommand \EOS [0]{\spacefactor3000\relax}%
\providecommand \BibitemShut  [1]{\csname bibitem#1\endcsname}%
\let\auto@bib@innerbib\@empty
\bibitem [{\citenamefont {Qi}\ and\ \citenamefont {Zhang}(2011)}]{QiXL11rmp}%
  \BibitemOpen
  \bibfield  {author} {\bibinfo {author} {\bibfnamefont {X.-L.}\ \bibnamefont
  {Qi}}\ and\ \bibinfo {author} {\bibfnamefont {S.-C.}\ \bibnamefont {Zhang}},\
  }\href {\doibase 10.1103/RevModPhys.83.1057} {\bibfield  {journal} {\bibinfo
  {journal} {Rev. Mod. Phys.}\ }\textbf {\bibinfo {volume} {83}},\ \bibinfo
  {pages} {1057} (\bibinfo {year} {2011})}\BibitemShut {NoStop}%
\bibitem [{\citenamefont {Hasan}\ and\ \citenamefont {Kane}(2010)}]{Kane10rmp}%
  \BibitemOpen
  \bibfield  {author} {\bibinfo {author} {\bibfnamefont {M.~Z.}\ \bibnamefont
  {Hasan}}\ and\ \bibinfo {author} {\bibfnamefont {C.~L.}\ \bibnamefont
  {Kane}},\ }\href {\doibase 10.1103/RevModPhys.82.3045} {\bibfield  {journal}
  {\bibinfo  {journal} {Rev. Mod. Phys.}\ }\textbf {\bibinfo {volume} {82}},\
  \bibinfo {pages} {3045} (\bibinfo {year} {2010})}\BibitemShut {NoStop}%
\bibitem [{\citenamefont {Su}\ \emph {et~al.}(1979)\citenamefont {Su},
  \citenamefont {Schrieffer},\ and\ \citenamefont {Heeger}}]{SSH79prl}%
  \BibitemOpen
  \bibfield  {author} {\bibinfo {author} {\bibfnamefont {W.~P.}\ \bibnamefont
  {Su}}, \bibinfo {author} {\bibfnamefont {J.~R.}\ \bibnamefont {Schrieffer}},
  \ and\ \bibinfo {author} {\bibfnamefont {A.~J.}\ \bibnamefont {Heeger}},\
  }\href {\doibase 10.1103/PhysRevLett.42.1698} {\bibfield  {journal} {\bibinfo
   {journal} {Phys. Rev. Lett.}\ }\textbf {\bibinfo {volume} {42}},\ \bibinfo
  {pages} {1698} (\bibinfo {year} {1979})}\BibitemShut {NoStop}%
\bibitem [{\citenamefont {King-Smith}\ and\ \citenamefont
  {Vanderbilt}(1993)}]{King93prb}%
  \BibitemOpen
  \bibfield  {author} {\bibinfo {author} {\bibfnamefont {R.~D.}\ \bibnamefont
  {King-Smith}}\ and\ \bibinfo {author} {\bibfnamefont {D.}~\bibnamefont
  {Vanderbilt}},\ }\href {\doibase 10.1103/PhysRevB.47.1651} {\bibfield
  {journal} {\bibinfo  {journal} {Phys. Rev. B}\ }\textbf {\bibinfo {volume}
  {47}},\ \bibinfo {pages} {1651} (\bibinfo {year} {1993})}\BibitemShut
  {NoStop}%
\bibitem [{\citenamefont {Shen}(2017)}]{SQS}%
  \BibitemOpen
  \bibfield  {author} {\bibinfo {author} {\bibfnamefont {S.-Q.}\ \bibnamefont
  {Shen}},\ }\href@noop {} {\emph {\bibinfo {title} {Topological Insulators:
  Dirac Equation in Condensed Matter}}},\ \bibinfo {edition} {2nd}\ ed.\
  (\bibinfo  {publisher} {Springer, Singapore},\ \bibinfo {year}
  {2017})\BibitemShut {NoStop}%
\bibitem [{\citenamefont {Bernevig}\ and\ \citenamefont
  {Hughes}(2013)}]{BernevigBook}%
  \BibitemOpen
  \bibfield  {author} {\bibinfo {author} {\bibfnamefont {B.~A.}\ \bibnamefont
  {Bernevig}}\ and\ \bibinfo {author} {\bibfnamefont {T.~L.}\ \bibnamefont
  {Hughes}},\ }\href@noop {} {\emph {\bibinfo {title} {Topological insulators
  and topological superconductors}}}\ (\bibinfo  {publisher} {Princeton
  University Press},\ \bibinfo {year} {2013})\BibitemShut {NoStop}%
\bibitem [{\citenamefont {Liu}\ and\ \citenamefont
  {Wakabayashi}(2017)}]{LiuF17prl}%
  \BibitemOpen
  \bibfield  {author} {\bibinfo {author} {\bibfnamefont {F.}~\bibnamefont
  {Liu}}\ and\ \bibinfo {author} {\bibfnamefont {K.}~\bibnamefont
  {Wakabayashi}},\ }\href {\doibase 10.1103/PhysRevLett.118.076803} {\bibfield
  {journal} {\bibinfo  {journal} {Phys. Rev. Lett.}\ }\textbf {\bibinfo
  {volume} {118}},\ \bibinfo {pages} {076803} (\bibinfo {year}
  {2017})}\BibitemShut {NoStop}%
\bibitem [{\citenamefont {Benalcazar}\ \emph
  {et~al.}(2017{\natexlab{a}})\citenamefont {Benalcazar}, \citenamefont
  {Bernevig},\ and\ \citenamefont {Hughes}}]{Benalcazar17Science}%
  \BibitemOpen
  \bibfield  {author} {\bibinfo {author} {\bibfnamefont {W.~A.}\ \bibnamefont
  {Benalcazar}}, \bibinfo {author} {\bibfnamefont {B.~A.}\ \bibnamefont
  {Bernevig}}, \ and\ \bibinfo {author} {\bibfnamefont {T.~L.}\ \bibnamefont
  {Hughes}},\ }\href {\doibase 10.1126/science.aah6442} {\bibfield  {journal}
  {\bibinfo  {journal} {Science}\ }\textbf {\bibinfo {volume} {357}},\ \bibinfo
  {pages} {61} (\bibinfo {year} {2017}{\natexlab{a}})}\BibitemShut {NoStop}%
\bibitem [{\citenamefont {Benalcazar}\ \emph
  {et~al.}(2017{\natexlab{b}})\citenamefont {Benalcazar}, \citenamefont
  {Bernevig},\ and\ \citenamefont {Hughes}}]{BBH17prb}%
  \BibitemOpen
  \bibfield  {author} {\bibinfo {author} {\bibfnamefont {W.~A.}\ \bibnamefont
  {Benalcazar}}, \bibinfo {author} {\bibfnamefont {B.~A.}\ \bibnamefont
  {Bernevig}}, \ and\ \bibinfo {author} {\bibfnamefont {T.~L.}\ \bibnamefont
  {Hughes}},\ }\href {\doibase 10.1103/PhysRevB.96.245115} {\bibfield
  {journal} {\bibinfo  {journal} {Phys. Rev. B}\ }\textbf {\bibinfo {volume}
  {96}},\ \bibinfo {pages} {245115} (\bibinfo {year}
  {2017}{\natexlab{b}})}\BibitemShut {NoStop}%
\bibitem [{\citenamefont {Langbehn}\ \emph {et~al.}(2017)\citenamefont
  {Langbehn}, \citenamefont {Peng}, \citenamefont {Trifunovic}, \citenamefont
  {von Oppen},\ and\ \citenamefont {Brouwer}}]{Langbehn17prl}%
  \BibitemOpen
  \bibfield  {author} {\bibinfo {author} {\bibfnamefont {J.}~\bibnamefont
  {Langbehn}}, \bibinfo {author} {\bibfnamefont {Y.}~\bibnamefont {Peng}},
  \bibinfo {author} {\bibfnamefont {L.}~\bibnamefont {Trifunovic}}, \bibinfo
  {author} {\bibfnamefont {F.}~\bibnamefont {von Oppen}}, \ and\ \bibinfo
  {author} {\bibfnamefont {P.~W.}\ \bibnamefont {Brouwer}},\ }\href {\doibase
  10.1103/PhysRevLett.119.246401} {\bibfield  {journal} {\bibinfo  {journal}
  {Phys. Rev. Lett.}\ }\textbf {\bibinfo {volume} {119}},\ \bibinfo {pages}
  {246401} (\bibinfo {year} {2017})}\BibitemShut {NoStop}%
\bibitem [{\citenamefont {Song}\ \emph {et~al.}(2017)\citenamefont {Song},
  \citenamefont {Fang},\ and\ \citenamefont {Fang}}]{SongZD17prl}%
  \BibitemOpen
  \bibfield  {author} {\bibinfo {author} {\bibfnamefont {Z.}~\bibnamefont
  {Song}}, \bibinfo {author} {\bibfnamefont {Z.}~\bibnamefont {Fang}}, \ and\
  \bibinfo {author} {\bibfnamefont {C.}~\bibnamefont {Fang}},\ }\href {\doibase
  10.1103/PhysRevLett.119.246402} {\bibfield  {journal} {\bibinfo  {journal}
  {Phys. Rev. Lett.}\ }\textbf {\bibinfo {volume} {119}},\ \bibinfo {pages}
  {246402} (\bibinfo {year} {2017})}\BibitemShut {NoStop}%
\bibitem [{\citenamefont {Geier}\ \emph {et~al.}(2018)\citenamefont {Geier},
  \citenamefont {Trifunovic}, \citenamefont {Hoskam},\ and\ \citenamefont
  {Brouwer}}]{Geier18prb}%
  \BibitemOpen
  \bibfield  {author} {\bibinfo {author} {\bibfnamefont {M.}~\bibnamefont
  {Geier}}, \bibinfo {author} {\bibfnamefont {L.}~\bibnamefont {Trifunovic}},
  \bibinfo {author} {\bibfnamefont {M.}~\bibnamefont {Hoskam}}, \ and\ \bibinfo
  {author} {\bibfnamefont {P.~W.}\ \bibnamefont {Brouwer}},\ }\href {\doibase
  10.1103/PhysRevB.97.205135} {\bibfield  {journal} {\bibinfo  {journal} {Phys.
  Rev. B}\ }\textbf {\bibinfo {volume} {97}},\ \bibinfo {pages} {205135}
  (\bibinfo {year} {2018})}\BibitemShut {NoStop}%
\bibitem [{\citenamefont {Schindler}\ \emph {et~al.}(2018)\citenamefont
  {Schindler}, \citenamefont {Wang}, \citenamefont {Vergniory}, \citenamefont
  {Cook}, \citenamefont {Murani}, \citenamefont {Sengupta}, \citenamefont
  {Kasumov}, \citenamefont {Deblock}, \citenamefont {Jeon}, \citenamefont
  {Drozdov}, \citenamefont {Bouchiat}, \citenamefont {Gu{\'e}ron},
  \citenamefont {Yazdani}, \citenamefont {Bernevig},\ and\ \citenamefont
  {Neupert}}]{Schindler18NP}%
  \BibitemOpen
  \bibfield  {author} {\bibinfo {author} {\bibfnamefont {F.}~\bibnamefont
  {Schindler}}, \bibinfo {author} {\bibfnamefont {Z.}~\bibnamefont {Wang}},
  \bibinfo {author} {\bibfnamefont {M.~G.}\ \bibnamefont {Vergniory}}, \bibinfo
  {author} {\bibfnamefont {A.~M.}\ \bibnamefont {Cook}}, \bibinfo {author}
  {\bibfnamefont {A.}~\bibnamefont {Murani}}, \bibinfo {author} {\bibfnamefont
  {S.}~\bibnamefont {Sengupta}}, \bibinfo {author} {\bibfnamefont {A.~Y.}\
  \bibnamefont {Kasumov}}, \bibinfo {author} {\bibfnamefont {R.}~\bibnamefont
  {Deblock}}, \bibinfo {author} {\bibfnamefont {S.}~\bibnamefont {Jeon}},
  \bibinfo {author} {\bibfnamefont {I.}~\bibnamefont {Drozdov}}, \bibinfo
  {author} {\bibfnamefont {H.}~\bibnamefont {Bouchiat}}, \bibinfo {author}
  {\bibfnamefont {S.}~\bibnamefont {Gu{\'e}ron}}, \bibinfo {author}
  {\bibfnamefont {A.}~\bibnamefont {Yazdani}}, \bibinfo {author} {\bibfnamefont
  {B.~A.}\ \bibnamefont {Bernevig}}, \ and\ \bibinfo {author} {\bibfnamefont
  {T.}~\bibnamefont {Neupert}},\ }\href {\doibase 10.1038/s41567-018-0224-7}
  {\bibfield  {journal} {\bibinfo  {journal} {Nat. Phys.}\ }\textbf {\bibinfo
  {volume} {14}},\ \bibinfo {pages} {918} (\bibinfo {year} {2018})}\BibitemShut
  {NoStop}%
\bibitem [{\citenamefont {Serra-Garcia}\ \emph {et~al.}(2018)\citenamefont
  {Serra-Garcia}, \citenamefont {Peri}, \citenamefont {S{\"u}sstrunk},
  \citenamefont {Bilal}, \citenamefont {Larsen}, \citenamefont {Villanueva},\
  and\ \citenamefont {Huber}}]{Serra-Garcia18nature}%
  \BibitemOpen
  \bibfield  {author} {\bibinfo {author} {\bibfnamefont {M.}~\bibnamefont
  {Serra-Garcia}}, \bibinfo {author} {\bibfnamefont {V.}~\bibnamefont {Peri}},
  \bibinfo {author} {\bibfnamefont {R.}~\bibnamefont {S{\"u}sstrunk}}, \bibinfo
  {author} {\bibfnamefont {O.~R.}\ \bibnamefont {Bilal}}, \bibinfo {author}
  {\bibfnamefont {T.}~\bibnamefont {Larsen}}, \bibinfo {author} {\bibfnamefont
  {L.~G.}\ \bibnamefont {Villanueva}}, \ and\ \bibinfo {author} {\bibfnamefont
  {S.~D.}\ \bibnamefont {Huber}},\ }\href {\doibase 10.1038/nature25156}
  {\bibfield  {journal} {\bibinfo  {journal} {Nature}\ }\textbf {\bibinfo
  {volume} {555}},\ \bibinfo {pages} {342} (\bibinfo {year}
  {2018})}\BibitemShut {NoStop}%
\bibitem [{\citenamefont {Peterson}\ \emph {et~al.}(2018)\citenamefont
  {Peterson}, \citenamefont {Benalcazar}, \citenamefont {Hughes},\ and\
  \citenamefont {Bahl}}]{Peterson18nature}%
  \BibitemOpen
  \bibfield  {author} {\bibinfo {author} {\bibfnamefont {C.~W.}\ \bibnamefont
  {Peterson}}, \bibinfo {author} {\bibfnamefont {W.~A.}\ \bibnamefont
  {Benalcazar}}, \bibinfo {author} {\bibfnamefont {T.~L.}\ \bibnamefont
  {Hughes}}, \ and\ \bibinfo {author} {\bibfnamefont {G.}~\bibnamefont
  {Bahl}},\ }\href {\doibase 10.1038/nature25777} {\bibfield  {journal}
  {\bibinfo  {journal} {Nature}\ }\textbf {\bibinfo {volume} {555}},\ \bibinfo
  {pages} {346} (\bibinfo {year} {2018})}\BibitemShut {NoStop}%
\bibitem [{\citenamefont {Ezawa}(2018)}]{Ezawa18prl}%
  \BibitemOpen
  \bibfield  {author} {\bibinfo {author} {\bibfnamefont {M.}~\bibnamefont
  {Ezawa}},\ }\href {\doibase 10.1103/PhysRevLett.120.026801} {\bibfield
  {journal} {\bibinfo  {journal} {Phys. Rev. Lett.}\ }\textbf {\bibinfo
  {volume} {120}},\ \bibinfo {pages} {026801} (\bibinfo {year}
  {2018})}\BibitemShut {NoStop}%
\bibitem [{\citenamefont {Kudo}\ \emph {et~al.}(2019)\citenamefont {Kudo},
  \citenamefont {Yoshida},\ and\ \citenamefont {Hatsugai}}]{Kudo19prl}%
  \BibitemOpen
  \bibfield  {author} {\bibinfo {author} {\bibfnamefont {K.}~\bibnamefont
  {Kudo}}, \bibinfo {author} {\bibfnamefont {T.}~\bibnamefont {Yoshida}}, \
  and\ \bibinfo {author} {\bibfnamefont {Y.}~\bibnamefont {Hatsugai}},\ }\href
  {\doibase 10.1103/PhysRevLett.123.196402} {\bibfield  {journal} {\bibinfo
  {journal} {Phys. Rev. Lett.}\ }\textbf {\bibinfo {volume} {123}},\ \bibinfo
  {pages} {196402} (\bibinfo {year} {2019})}\BibitemShut {NoStop}%
\bibitem [{\citenamefont {Volpez}\ \emph {et~al.}(2019)\citenamefont {Volpez},
  \citenamefont {Loss},\ and\ \citenamefont {Klinovaja}}]{Volpez19prl}%
  \BibitemOpen
  \bibfield  {author} {\bibinfo {author} {\bibfnamefont {Y.}~\bibnamefont
  {Volpez}}, \bibinfo {author} {\bibfnamefont {D.}~\bibnamefont {Loss}}, \ and\
  \bibinfo {author} {\bibfnamefont {J.}~\bibnamefont {Klinovaja}},\ }\href
  {\doibase 10.1103/PhysRevLett.122.126402} {\bibfield  {journal} {\bibinfo
  {journal} {Phys. Rev. Lett.}\ }\textbf {\bibinfo {volume} {122}},\ \bibinfo
  {pages} {126402} (\bibinfo {year} {2019})}\BibitemShut {NoStop}%
\bibitem [{\citenamefont {Sheng}\ \emph {et~al.}(2019)\citenamefont {Sheng},
  \citenamefont {Chen}, \citenamefont {Liu}, \citenamefont {Chen},
  \citenamefont {Yu}, \citenamefont {Zhao},\ and\ \citenamefont
  {Yang}}]{XLSheng19prl}%
  \BibitemOpen
  \bibfield  {author} {\bibinfo {author} {\bibfnamefont {X.-L.}\ \bibnamefont
  {Sheng}}, \bibinfo {author} {\bibfnamefont {C.}~\bibnamefont {Chen}},
  \bibinfo {author} {\bibfnamefont {H.}~\bibnamefont {Liu}}, \bibinfo {author}
  {\bibfnamefont {Z.}~\bibnamefont {Chen}}, \bibinfo {author} {\bibfnamefont
  {Z.-M.}\ \bibnamefont {Yu}}, \bibinfo {author} {\bibfnamefont {Y.~X.}\
  \bibnamefont {Zhao}}, \ and\ \bibinfo {author} {\bibfnamefont {S.~A.}\
  \bibnamefont {Yang}},\ }\href {\doibase 10.1103/PhysRevLett.123.256402}
  {\bibfield  {journal} {\bibinfo  {journal} {Phys. Rev. Lett.}\ }\textbf
  {\bibinfo {volume} {123}},\ \bibinfo {pages} {256402} (\bibinfo {year}
  {2019})}\BibitemShut {NoStop}%
\bibitem [{\citenamefont {Hua}\ \emph {et~al.}(2020)\citenamefont {Hua},
  \citenamefont {Chen}, \citenamefont {Zhou},\ and\ \citenamefont
  {Xu}}]{HuaCB20prb}%
  \BibitemOpen
  \bibfield  {author} {\bibinfo {author} {\bibfnamefont {C.-B.}\ \bibnamefont
  {Hua}}, \bibinfo {author} {\bibfnamefont {R.}~\bibnamefont {Chen}}, \bibinfo
  {author} {\bibfnamefont {B.}~\bibnamefont {Zhou}}, \ and\ \bibinfo {author}
  {\bibfnamefont {D.-H.}\ \bibnamefont {Xu}},\ }\href {\doibase
  10.1103/PhysRevB.102.241102} {\bibfield  {journal} {\bibinfo  {journal}
  {Phys. Rev. B}\ }\textbf {\bibinfo {volume} {102}},\ \bibinfo {pages}
  {241102} (\bibinfo {year} {2020})}\BibitemShut {NoStop}%
\bibitem [{\citenamefont {Chen}\ \emph {et~al.}(2020)\citenamefont {Chen},
  \citenamefont {Chen}, \citenamefont {Gao}, \citenamefont {Zhou},\ and\
  \citenamefont {Xu}}]{ChenR20prl}%
  \BibitemOpen
  \bibfield  {author} {\bibinfo {author} {\bibfnamefont {R.}~\bibnamefont
  {Chen}}, \bibinfo {author} {\bibfnamefont {C.-Z.}\ \bibnamefont {Chen}},
  \bibinfo {author} {\bibfnamefont {J.-H.}\ \bibnamefont {Gao}}, \bibinfo
  {author} {\bibfnamefont {B.}~\bibnamefont {Zhou}}, \ and\ \bibinfo {author}
  {\bibfnamefont {D.-H.}\ \bibnamefont {Xu}},\ }\href {\doibase
  10.1103/PhysRevLett.124.036803} {\bibfield  {journal} {\bibinfo  {journal}
  {Phys. Rev. Lett.}\ }\textbf {\bibinfo {volume} {124}},\ \bibinfo {pages}
  {036803} (\bibinfo {year} {2020})}\BibitemShut {NoStop}%
\bibitem [{\citenamefont {Zhang}\ \emph
  {et~al.}(2020{\natexlab{a}})\citenamefont {Zhang}, \citenamefont {Wu},\ and\
  \citenamefont {Das~Sarma}}]{ZhangRX20prl}%
  \BibitemOpen
  \bibfield  {author} {\bibinfo {author} {\bibfnamefont {R.-X.}\ \bibnamefont
  {Zhang}}, \bibinfo {author} {\bibfnamefont {F.}~\bibnamefont {Wu}}, \ and\
  \bibinfo {author} {\bibfnamefont {S.}~\bibnamefont {Das~Sarma}},\ }\href
  {\doibase 10.1103/PhysRevLett.124.136407} {\bibfield  {journal} {\bibinfo
  {journal} {Phys. Rev. Lett.}\ }\textbf {\bibinfo {volume} {124}},\ \bibinfo
  {pages} {136407} (\bibinfo {year} {2020}{\natexlab{a}})}\BibitemShut
  {NoStop}%
\bibitem [{\citenamefont {Roy}(2019)}]{Roy19prr}%
  \BibitemOpen
  \bibfield  {author} {\bibinfo {author} {\bibfnamefont {B.}~\bibnamefont
  {Roy}},\ }\href {\doibase 10.1103/PhysRevResearch.1.032048} {\bibfield
  {journal} {\bibinfo  {journal} {Phys. Rev. Research}\ }\textbf {\bibinfo
  {volume} {1}},\ \bibinfo {pages} {032048} (\bibinfo {year}
  {2019})}\BibitemShut {NoStop}%
\bibitem [{\citenamefont {Peng}\ and\ \citenamefont
  {Refael}(2019)}]{PengY19prl}%
  \BibitemOpen
  \bibfield  {author} {\bibinfo {author} {\bibfnamefont {Y.}~\bibnamefont
  {Peng}}\ and\ \bibinfo {author} {\bibfnamefont {G.}~\bibnamefont {Refael}},\
  }\href {\doibase 10.1103/PhysRevLett.123.016806} {\bibfield  {journal}
  {\bibinfo  {journal} {Phys. Rev. Lett.}\ }\textbf {\bibinfo {volume} {123}},\
  \bibinfo {pages} {016806} (\bibinfo {year} {2019})}\BibitemShut {NoStop}%
\bibitem [{\citenamefont {Li}\ and\ \citenamefont {Wu}(2020)}]{LiC20prb}%
  \BibitemOpen
  \bibfield  {author} {\bibinfo {author} {\bibfnamefont {C.-A.}\ \bibnamefont
  {Li}}\ and\ \bibinfo {author} {\bibfnamefont {S.-S.}\ \bibnamefont {Wu}},\
  }\href {\doibase 10.1103/PhysRevB.101.195309} {\bibfield  {journal} {\bibinfo
   {journal} {Phys. Rev. B}\ }\textbf {\bibinfo {volume} {101}},\ \bibinfo
  {pages} {195309} (\bibinfo {year} {2020})}\BibitemShut {NoStop}%
\bibitem [{\citenamefont {Li}\ \emph {et~al.}(2020)\citenamefont {Li},
  \citenamefont {Fu}, \citenamefont {Hu}, \citenamefont {Li},\ and\
  \citenamefont {Shen}}]{LiCA20prl}%
  \BibitemOpen
  \bibfield  {author} {\bibinfo {author} {\bibfnamefont {C.-A.}\ \bibnamefont
  {Li}}, \bibinfo {author} {\bibfnamefont {B.}~\bibnamefont {Fu}}, \bibinfo
  {author} {\bibfnamefont {Z.-A.}\ \bibnamefont {Hu}}, \bibinfo {author}
  {\bibfnamefont {J.}~\bibnamefont {Li}}, \ and\ \bibinfo {author}
  {\bibfnamefont {S.-Q.}\ \bibnamefont {Shen}},\ }\href {\doibase
  10.1103/PhysRevLett.125.166801} {\bibfield  {journal} {\bibinfo  {journal}
  {Phys. Rev. Lett.}\ }\textbf {\bibinfo {volume} {125}},\ \bibinfo {pages}
  {166801} (\bibinfo {year} {2020})}\BibitemShut {NoStop}%
\bibitem [{\citenamefont {Li}\ \emph {et~al.}(2021{\natexlab{a}})\citenamefont
  {Li}, \citenamefont {Zhang}, \citenamefont {Li},\ and\ \citenamefont
  {Trauzettel}}]{LiCA21prl}%
  \BibitemOpen
  \bibfield  {author} {\bibinfo {author} {\bibfnamefont {C.-A.}\ \bibnamefont
  {Li}}, \bibinfo {author} {\bibfnamefont {S.-B.}\ \bibnamefont {Zhang}},
  \bibinfo {author} {\bibfnamefont {J.}~\bibnamefont {Li}}, \ and\ \bibinfo
  {author} {\bibfnamefont {B.}~\bibnamefont {Trauzettel}},\ }\href {\doibase
  10.1103/PhysRevLett.127.026803} {\bibfield  {journal} {\bibinfo  {journal}
  {Phys. Rev. Lett.}\ }\textbf {\bibinfo {volume} {127}},\ \bibinfo {pages}
  {026803} (\bibinfo {year} {2021}{\natexlab{a}})}\BibitemShut {NoStop}%
\bibitem [{\citenamefont {Chen}\ \emph {et~al.}(2019)\citenamefont {Chen},
  \citenamefont {Deng}, \citenamefont {Shi}, \citenamefont {Zhao},
  \citenamefont {Chen},\ and\ \citenamefont {Dong}}]{ChenXD19prl}%
  \BibitemOpen
  \bibfield  {author} {\bibinfo {author} {\bibfnamefont {X.-D.}\ \bibnamefont
  {Chen}}, \bibinfo {author} {\bibfnamefont {W.-M.}\ \bibnamefont {Deng}},
  \bibinfo {author} {\bibfnamefont {F.-L.}\ \bibnamefont {Shi}}, \bibinfo
  {author} {\bibfnamefont {F.-L.}\ \bibnamefont {Zhao}}, \bibinfo {author}
  {\bibfnamefont {M.}~\bibnamefont {Chen}}, \ and\ \bibinfo {author}
  {\bibfnamefont {J.-W.}\ \bibnamefont {Dong}},\ }\href {\doibase
  10.1103/PhysRevLett.122.233902} {\bibfield  {journal} {\bibinfo  {journal}
  {Phys. Rev. Lett.}\ }\textbf {\bibinfo {volume} {122}},\ \bibinfo {pages}
  {233902} (\bibinfo {year} {2019})}\BibitemShut {NoStop}%
\bibitem [{\citenamefont {Qi}\ \emph {et~al.}(2020)\citenamefont {Qi},
  \citenamefont {Qiu}, \citenamefont {Xiao}, \citenamefont {He}, \citenamefont
  {Ke},\ and\ \citenamefont {Liu}}]{QiY20prl}%
  \BibitemOpen
  \bibfield  {author} {\bibinfo {author} {\bibfnamefont {Y.}~\bibnamefont
  {Qi}}, \bibinfo {author} {\bibfnamefont {C.}~\bibnamefont {Qiu}}, \bibinfo
  {author} {\bibfnamefont {M.}~\bibnamefont {Xiao}}, \bibinfo {author}
  {\bibfnamefont {H.}~\bibnamefont {He}}, \bibinfo {author} {\bibfnamefont
  {M.}~\bibnamefont {Ke}}, \ and\ \bibinfo {author} {\bibfnamefont
  {Z.}~\bibnamefont {Liu}},\ }\href {\doibase 10.1103/PhysRevLett.124.206601}
  {\bibfield  {journal} {\bibinfo  {journal} {Phys. Rev. Lett.}\ }\textbf
  {\bibinfo {volume} {124}},\ \bibinfo {pages} {206601} (\bibinfo {year}
  {2020})}\BibitemShut {NoStop}%
\bibitem [{\citenamefont {Wei}\ \emph {et~al.}(2021)\citenamefont {Wei},
  \citenamefont {Zhang}, \citenamefont {Deng}, \citenamefont {Lu},
  \citenamefont {Huang}, \citenamefont {Yan}, \citenamefont {Chen},
  \citenamefont {Liu},\ and\ \citenamefont {Jia}}]{WeiQ21prl}%
  \BibitemOpen
  \bibfield  {author} {\bibinfo {author} {\bibfnamefont {Q.}~\bibnamefont
  {Wei}}, \bibinfo {author} {\bibfnamefont {X.}~\bibnamefont {Zhang}}, \bibinfo
  {author} {\bibfnamefont {W.}~\bibnamefont {Deng}}, \bibinfo {author}
  {\bibfnamefont {J.}~\bibnamefont {Lu}}, \bibinfo {author} {\bibfnamefont
  {X.}~\bibnamefont {Huang}}, \bibinfo {author} {\bibfnamefont
  {M.}~\bibnamefont {Yan}}, \bibinfo {author} {\bibfnamefont {G.}~\bibnamefont
  {Chen}}, \bibinfo {author} {\bibfnamefont {Z.}~\bibnamefont {Liu}}, \ and\
  \bibinfo {author} {\bibfnamefont {S.}~\bibnamefont {Jia}},\ }\href {\doibase
  10.1103/PhysRevLett.127.255501} {\bibfield  {journal} {\bibinfo  {journal}
  {Phys. Rev. Lett.}\ }\textbf {\bibinfo {volume} {127}},\ \bibinfo {pages}
  {255501} (\bibinfo {year} {2021})}\BibitemShut {NoStop}%
\bibitem [{\citenamefont {Ning}\ \emph {et~al.}(2022)\citenamefont {Ning},
  \citenamefont {Fu}, \citenamefont {Xu},\ and\ \citenamefont
  {Wang}}]{NingZ22arxiv}%
  \BibitemOpen
  \bibfield  {author} {\bibinfo {author} {\bibfnamefont {Z.}~\bibnamefont
  {Ning}}, \bibinfo {author} {\bibfnamefont {B.}~\bibnamefont {Fu}}, \bibinfo
  {author} {\bibfnamefont {D.-H.}\ \bibnamefont {Xu}}, \ and\ \bibinfo {author}
  {\bibfnamefont {R.}~\bibnamefont {Wang}},\ }\href@noop {} {\enquote {\bibinfo
  {title} {Tailoring quadrupole topological insulators with periodic driving
  and disorder},}\ } (\bibinfo {year} {2022}),\ \Eprint
  {http://arxiv.org/abs/2201.02414} {arXiv:2201.02414 [cond-mat.mes-hall]}
  \BibitemShut {NoStop}%
\bibitem [{\citenamefont {Wang}\ \emph {et~al.}(2018)\citenamefont {Wang},
  \citenamefont {Liu}, \citenamefont {Lu},\ and\ \citenamefont
  {Zhang}}]{WangQ18prl}%
  \BibitemOpen
  \bibfield  {author} {\bibinfo {author} {\bibfnamefont {Q.}~\bibnamefont
  {Wang}}, \bibinfo {author} {\bibfnamefont {C.-C.}\ \bibnamefont {Liu}},
  \bibinfo {author} {\bibfnamefont {Y.-M.}\ \bibnamefont {Lu}}, \ and\ \bibinfo
  {author} {\bibfnamefont {F.}~\bibnamefont {Zhang}},\ }\href {\doibase
  10.1103/PhysRevLett.121.186801} {\bibfield  {journal} {\bibinfo  {journal}
  {Phys. Rev. Lett.}\ }\textbf {\bibinfo {volume} {121}},\ \bibinfo {pages}
  {186801} (\bibinfo {year} {2018})}\BibitemShut {NoStop}%
\bibitem [{\citenamefont {Yan}\ \emph {et~al.}(2018)\citenamefont {Yan},
  \citenamefont {Song},\ and\ \citenamefont {Wang}}]{YanZB18prl}%
  \BibitemOpen
  \bibfield  {author} {\bibinfo {author} {\bibfnamefont {Z.}~\bibnamefont
  {Yan}}, \bibinfo {author} {\bibfnamefont {F.}~\bibnamefont {Song}}, \ and\
  \bibinfo {author} {\bibfnamefont {Z.}~\bibnamefont {Wang}},\ }\href {\doibase
  10.1103/PhysRevLett.121.096803} {\bibfield  {journal} {\bibinfo  {journal}
  {Phys. Rev. Lett.}\ }\textbf {\bibinfo {volume} {121}},\ \bibinfo {pages}
  {096803} (\bibinfo {year} {2018})}\BibitemShut {NoStop}%
\bibitem [{\citenamefont {Zhang}\ \emph {et~al.}(2019)\citenamefont {Zhang},
  \citenamefont {Cole}, \citenamefont {Wu},\ and\ \citenamefont
  {Das~Sarma}}]{ZhangRX19prl}%
  \BibitemOpen
  \bibfield  {author} {\bibinfo {author} {\bibfnamefont {R.-X.}\ \bibnamefont
  {Zhang}}, \bibinfo {author} {\bibfnamefont {W.~S.}\ \bibnamefont {Cole}},
  \bibinfo {author} {\bibfnamefont {X.}~\bibnamefont {Wu}}, \ and\ \bibinfo
  {author} {\bibfnamefont {S.}~\bibnamefont {Das~Sarma}},\ }\href {\doibase
  10.1103/PhysRevLett.123.167001} {\bibfield  {journal} {\bibinfo  {journal}
  {Phys. Rev. Lett.}\ }\textbf {\bibinfo {volume} {123}},\ \bibinfo {pages}
  {167001} (\bibinfo {year} {2019})}\BibitemShut {NoStop}%
\bibitem [{\citenamefont {Zhang}\ and\ \citenamefont
  {Trauzettel}(2020)}]{ZhangSB20prr}%
  \BibitemOpen
  \bibfield  {author} {\bibinfo {author} {\bibfnamefont {S.-B.}\ \bibnamefont
  {Zhang}}\ and\ \bibinfo {author} {\bibfnamefont {B.}~\bibnamefont
  {Trauzettel}},\ }\href {\doibase 10.1103/PhysRevResearch.2.012018} {\bibfield
   {journal} {\bibinfo  {journal} {Phys. Rev. Research}\ }\textbf {\bibinfo
  {volume} {2}},\ \bibinfo {pages} {012018} (\bibinfo {year}
  {2020})}\BibitemShut {NoStop}%
\bibitem [{\citenamefont {Zhu}(2018)}]{XYZhu18prb}%
  \BibitemOpen
  \bibfield  {author} {\bibinfo {author} {\bibfnamefont {X.}~\bibnamefont
  {Zhu}},\ }\href {\doibase 10.1103/PhysRevB.97.205134} {\bibfield  {journal}
  {\bibinfo  {journal} {Phys. Rev. B}\ }\textbf {\bibinfo {volume} {97}},\
  \bibinfo {pages} {205134} (\bibinfo {year} {2018})}\BibitemShut {NoStop}%
\bibitem [{\citenamefont {Wu}\ \emph {et~al.}(2020)\citenamefont {Wu},
  \citenamefont {Hou}, \citenamefont {Li}, \citenamefont {Luo}, \citenamefont
  {Shi},\ and\ \citenamefont {Zhang}}]{WuYJ20prl}%
  \BibitemOpen
  \bibfield  {author} {\bibinfo {author} {\bibfnamefont {Y.-J.}\ \bibnamefont
  {Wu}}, \bibinfo {author} {\bibfnamefont {J.}~\bibnamefont {Hou}}, \bibinfo
  {author} {\bibfnamefont {Y.-M.}\ \bibnamefont {Li}}, \bibinfo {author}
  {\bibfnamefont {X.-W.}\ \bibnamefont {Luo}}, \bibinfo {author} {\bibfnamefont
  {X.}~\bibnamefont {Shi}}, \ and\ \bibinfo {author} {\bibfnamefont
  {C.}~\bibnamefont {Zhang}},\ }\href {\doibase 10.1103/PhysRevLett.124.227001}
  {\bibfield  {journal} {\bibinfo  {journal} {Phys. Rev. Lett.}\ }\textbf
  {\bibinfo {volume} {124}},\ \bibinfo {pages} {227001} (\bibinfo {year}
  {2020})}\BibitemShut {NoStop}%
\bibitem [{\citenamefont {Zhang}\ \emph
  {et~al.}(2020{\natexlab{b}})\citenamefont {Zhang}, \citenamefont {Rui},
  \citenamefont {Calzona}, \citenamefont {Choi}, \citenamefont {Schnyder},\
  and\ \citenamefont {Trauzettel}}]{ZhangSB20prr2}%
  \BibitemOpen
  \bibfield  {author} {\bibinfo {author} {\bibfnamefont {S.-B.}\ \bibnamefont
  {Zhang}}, \bibinfo {author} {\bibfnamefont {W.~B.}\ \bibnamefont {Rui}},
  \bibinfo {author} {\bibfnamefont {A.}~\bibnamefont {Calzona}}, \bibinfo
  {author} {\bibfnamefont {S.-J.}\ \bibnamefont {Choi}}, \bibinfo {author}
  {\bibfnamefont {A.~P.}\ \bibnamefont {Schnyder}}, \ and\ \bibinfo {author}
  {\bibfnamefont {B.}~\bibnamefont {Trauzettel}},\ }\href {\doibase
  10.1103/PhysRevResearch.2.043025} {\bibfield  {journal} {\bibinfo  {journal}
  {Phys. Rev. Research}\ }\textbf {\bibinfo {volume} {2}},\ \bibinfo {pages}
  {043025} (\bibinfo {year} {2020}{\natexlab{b}})}\BibitemShut {NoStop}%
\bibitem [{\citenamefont {Wang}\ \emph {et~al.}(2020)\citenamefont {Wang},
  \citenamefont {Lin}, \citenamefont {Jiang}, \citenamefont {Guo},\ and\
  \citenamefont {Jiang}}]{WangHX20prl}%
  \BibitemOpen
  \bibfield  {author} {\bibinfo {author} {\bibfnamefont {H.-X.}\ \bibnamefont
  {Wang}}, \bibinfo {author} {\bibfnamefont {Z.-K.}\ \bibnamefont {Lin}},
  \bibinfo {author} {\bibfnamefont {B.}~\bibnamefont {Jiang}}, \bibinfo
  {author} {\bibfnamefont {G.-Y.}\ \bibnamefont {Guo}}, \ and\ \bibinfo
  {author} {\bibfnamefont {J.-H.}\ \bibnamefont {Jiang}},\ }\href {\doibase
  10.1103/PhysRevLett.125.146401} {\bibfield  {journal} {\bibinfo  {journal}
  {Phys. Rev. Lett.}\ }\textbf {\bibinfo {volume} {125}},\ \bibinfo {pages}
  {146401} (\bibinfo {year} {2020})}\BibitemShut {NoStop}%
\bibitem [{\citenamefont {Ghorashi}\ \emph {et~al.}(2020)\citenamefont
  {Ghorashi}, \citenamefont {Li},\ and\ \citenamefont
  {Hughes}}]{Ghorashi20prl}%
  \BibitemOpen
  \bibfield  {author} {\bibinfo {author} {\bibfnamefont {S.~A.~A.}\
  \bibnamefont {Ghorashi}}, \bibinfo {author} {\bibfnamefont {T.}~\bibnamefont
  {Li}}, \ and\ \bibinfo {author} {\bibfnamefont {T.~L.}\ \bibnamefont
  {Hughes}},\ }\href {\doibase 10.1103/PhysRevLett.125.266804} {\bibfield
  {journal} {\bibinfo  {journal} {Phys. Rev. Lett.}\ }\textbf {\bibinfo
  {volume} {125}},\ \bibinfo {pages} {266804} (\bibinfo {year}
  {2020})}\BibitemShut {NoStop}%
\bibitem [{\citenamefont {Wang}\ \emph {et~al.}(2019)\citenamefont {Wang},
  \citenamefont {Wieder}, \citenamefont {Li}, \citenamefont {Yan},\ and\
  \citenamefont {Bernevig}}]{WangZJ19prl}%
  \BibitemOpen
  \bibfield  {author} {\bibinfo {author} {\bibfnamefont {Z.}~\bibnamefont
  {Wang}}, \bibinfo {author} {\bibfnamefont {B.~J.}\ \bibnamefont {Wieder}},
  \bibinfo {author} {\bibfnamefont {J.}~\bibnamefont {Li}}, \bibinfo {author}
  {\bibfnamefont {B.}~\bibnamefont {Yan}}, \ and\ \bibinfo {author}
  {\bibfnamefont {B.~A.}\ \bibnamefont {Bernevig}},\ }\href {\doibase
  10.1103/PhysRevLett.123.186401} {\bibfield  {journal} {\bibinfo  {journal}
  {Phys. Rev. Lett.}\ }\textbf {\bibinfo {volume} {123}},\ \bibinfo {pages}
  {186401} (\bibinfo {year} {2019})}\BibitemShut {NoStop}%
\bibitem [{\citenamefont {Li}\ \emph {et~al.}(2021{\natexlab{b}})\citenamefont
  {Li}, \citenamefont {Choi}, \citenamefont {Zhang},\ and\ \citenamefont
  {Trauzettel}}]{LICA22arxiv}%
  \BibitemOpen
  \bibfield  {author} {\bibinfo {author} {\bibfnamefont {C.-A.}\ \bibnamefont
  {Li}}, \bibinfo {author} {\bibfnamefont {S.-J.}\ \bibnamefont {Choi}},
  \bibinfo {author} {\bibfnamefont {S.-B.}\ \bibnamefont {Zhang}}, \ and\
  \bibinfo {author} {\bibfnamefont {B.}~\bibnamefont {Trauzettel}},\
  }\href@noop {} {} (\bibinfo {year} {2021}{\natexlab{b}}),\ \Eprint
  {http://arxiv.org/abs/2112.07697} {arXiv:2112.07697 [cond-mat.mes-hall]}
  \BibitemShut {NoStop}%
\bibitem [{\citenamefont {Schnyder}\ and\ \citenamefont
  {Ryu}(2011)}]{Schnyder11prb}%
  \BibitemOpen
  \bibfield  {author} {\bibinfo {author} {\bibfnamefont {A.~P.}\ \bibnamefont
  {Schnyder}}\ and\ \bibinfo {author} {\bibfnamefont {S.}~\bibnamefont {Ryu}},\
  }\href {\doibase 10.1103/PhysRevB.84.060504} {\bibfield  {journal} {\bibinfo
  {journal} {Phys. Rev. B}\ }\textbf {\bibinfo {volume} {84}},\ \bibinfo
  {pages} {060504} (\bibinfo {year} {2011})}\BibitemShut {NoStop}%
\bibitem [{\citenamefont {Heikkil{\"a}}\ \emph {et~al.}(2011)\citenamefont
  {Heikkil{\"a}}, \citenamefont {Kopnin},\ and\ \citenamefont
  {Volovik}}]{Heikkila11jetp}%
  \BibitemOpen
  \bibfield  {author} {\bibinfo {author} {\bibfnamefont {T.~T.}\ \bibnamefont
  {Heikkil{\"a}}}, \bibinfo {author} {\bibfnamefont {N.~B.}\ \bibnamefont
  {Kopnin}}, \ and\ \bibinfo {author} {\bibfnamefont {G.~E.}\ \bibnamefont
  {Volovik}},\ }\href {\doibase 10.1134/S0021364011150045} {\bibfield
  {journal} {\bibinfo  {journal} {JETP Letters}\ }\textbf {\bibinfo {volume}
  {94}},\ \bibinfo {pages} {233} (\bibinfo {year} {2011})}\BibitemShut
  {NoStop}%
\bibitem [{\citenamefont {Montambaux}\ \emph {et~al.}(2009)\citenamefont
  {Montambaux}, \citenamefont {Pi\'echon}, \citenamefont {Fuchs},\ and\
  \citenamefont {Goerbig}}]{Montambaux09prb}%
  \BibitemOpen
  \bibfield  {author} {\bibinfo {author} {\bibfnamefont {G.}~\bibnamefont
  {Montambaux}}, \bibinfo {author} {\bibfnamefont {F.}~\bibnamefont
  {Pi\'echon}}, \bibinfo {author} {\bibfnamefont {J.-N.}\ \bibnamefont
  {Fuchs}}, \ and\ \bibinfo {author} {\bibfnamefont {M.~O.}\ \bibnamefont
  {Goerbig}},\ }\href {\doibase 10.1103/PhysRevB.80.153412} {\bibfield
  {journal} {\bibinfo  {journal} {Phys. Rev. B}\ }\textbf {\bibinfo {volume}
  {80}},\ \bibinfo {pages} {153412} (\bibinfo {year} {2009})}\BibitemShut
  {NoStop}%
\bibitem [{\citenamefont {Alexandradinata}\ \emph {et~al.}(2014)\citenamefont
  {Alexandradinata}, \citenamefont {Dai},\ and\ \citenamefont
  {Bernevig}}]{Alexandradinata14prb}%
  \BibitemOpen
  \bibfield  {author} {\bibinfo {author} {\bibfnamefont {A.}~\bibnamefont
  {Alexandradinata}}, \bibinfo {author} {\bibfnamefont {X.}~\bibnamefont
  {Dai}}, \ and\ \bibinfo {author} {\bibfnamefont {B.~A.}\ \bibnamefont
  {Bernevig}},\ }\href {\doibase 10.1103/PhysRevB.89.155114} {\bibfield
  {journal} {\bibinfo  {journal} {Phys. Rev. B}\ }\textbf {\bibinfo {volume}
  {89}},\ \bibinfo {pages} {155114} (\bibinfo {year} {2014})}\BibitemShut
  {NoStop}%
\bibitem [{\citenamefont {Wang}\ \emph {et~al.}(2009)\citenamefont {Wang},
  \citenamefont {Chong}, \citenamefont {Joannopoulos},\ and\ \citenamefont
  {Solja{\v{c}}i{\'{c}}}}]{WangZ09nature}%
  \BibitemOpen
  \bibfield  {author} {\bibinfo {author} {\bibfnamefont {Z.}~\bibnamefont
  {Wang}}, \bibinfo {author} {\bibfnamefont {Y.}~\bibnamefont {Chong}},
  \bibinfo {author} {\bibfnamefont {J.~D.}\ \bibnamefont {Joannopoulos}}, \
  and\ \bibinfo {author} {\bibfnamefont {M.}~\bibnamefont
  {Solja{\v{c}}i{\'{c}}}},\ }\href {\doibase 10.1038/nature08293} {\bibfield
  {journal} {\bibinfo  {journal} {Nature}\ }\textbf {\bibinfo {volume} {461}},\
  \bibinfo {pages} {772} (\bibinfo {year} {2009})}\BibitemShut {NoStop}%
\bibitem [{\citenamefont {Xie}\ \emph {et~al.}(2019)\citenamefont {Xie},
  \citenamefont {Su}, \citenamefont {Wang}, \citenamefont {Su}, \citenamefont
  {Shen}, \citenamefont {Zhan}, \citenamefont {Lu}, \citenamefont {Wang},\ and\
  \citenamefont {Chen}}]{XieBY19prl}%
  \BibitemOpen
  \bibfield  {author} {\bibinfo {author} {\bibfnamefont {B.-Y.}\ \bibnamefont
  {Xie}}, \bibinfo {author} {\bibfnamefont {G.-X.}\ \bibnamefont {Su}},
  \bibinfo {author} {\bibfnamefont {H.-F.}\ \bibnamefont {Wang}}, \bibinfo
  {author} {\bibfnamefont {H.}~\bibnamefont {Su}}, \bibinfo {author}
  {\bibfnamefont {X.-P.}\ \bibnamefont {Shen}}, \bibinfo {author}
  {\bibfnamefont {P.}~\bibnamefont {Zhan}}, \bibinfo {author} {\bibfnamefont
  {M.-H.}\ \bibnamefont {Lu}}, \bibinfo {author} {\bibfnamefont {Z.-L.}\
  \bibnamefont {Wang}}, \ and\ \bibinfo {author} {\bibfnamefont {Y.-F.}\
  \bibnamefont {Chen}},\ }\href {\doibase 10.1103/PhysRevLett.122.233903}
  {\bibfield  {journal} {\bibinfo  {journal} {Phys. Rev. Lett.}\ }\textbf
  {\bibinfo {volume} {122}},\ \bibinfo {pages} {233903} (\bibinfo {year}
  {2019})}\BibitemShut {NoStop}%
\bibitem [{\citenamefont {Ni}\ \emph {et~al.}(2019)\citenamefont {Ni},
  \citenamefont {Weiner}, \citenamefont {Al{\`u}},\ and\ \citenamefont
  {Khanikaev}}]{Ni19nm}%
  \BibitemOpen
  \bibfield  {author} {\bibinfo {author} {\bibfnamefont {X.}~\bibnamefont
  {Ni}}, \bibinfo {author} {\bibfnamefont {M.}~\bibnamefont {Weiner}}, \bibinfo
  {author} {\bibfnamefont {A.}~\bibnamefont {Al{\`u}}}, \ and\ \bibinfo
  {author} {\bibfnamefont {A.~B.}\ \bibnamefont {Khanikaev}},\ }\href {\doibase
  10.1038/s41563-018-0252-9} {\bibfield  {journal} {\bibinfo  {journal} {Nat.
  Mater.}\ }\textbf {\bibinfo {volume} {18}},\ \bibinfo {pages} {113} (\bibinfo
  {year} {2019})}\BibitemShut {NoStop}%
\bibitem [{\citenamefont {Imhof}\ \emph {et~al.}(2018)\citenamefont {Imhof},
  \citenamefont {Berger}, \citenamefont {Bayer}, \citenamefont {Brehm},
  \citenamefont {Molenkamp}, \citenamefont {Kiessling}, \citenamefont
  {Schindler}, \citenamefont {Lee}, \citenamefont {Greiter}, \citenamefont
  {Neupert},\ and\ \citenamefont {Thomale}}]{Imhof18np}%
  \BibitemOpen
  \bibfield  {author} {\bibinfo {author} {\bibfnamefont {S.}~\bibnamefont
  {Imhof}}, \bibinfo {author} {\bibfnamefont {C.}~\bibnamefont {Berger}},
  \bibinfo {author} {\bibfnamefont {F.}~\bibnamefont {Bayer}}, \bibinfo
  {author} {\bibfnamefont {J.}~\bibnamefont {Brehm}}, \bibinfo {author}
  {\bibfnamefont {L.~W.}\ \bibnamefont {Molenkamp}}, \bibinfo {author}
  {\bibfnamefont {T.}~\bibnamefont {Kiessling}}, \bibinfo {author}
  {\bibfnamefont {F.}~\bibnamefont {Schindler}}, \bibinfo {author}
  {\bibfnamefont {C.~H.}\ \bibnamefont {Lee}}, \bibinfo {author} {\bibfnamefont
  {M.}~\bibnamefont {Greiter}}, \bibinfo {author} {\bibfnamefont
  {T.}~\bibnamefont {Neupert}}, \ and\ \bibinfo {author} {\bibfnamefont
  {R.}~\bibnamefont {Thomale}},\ }\href {\doibase 10.1038/s41567-018-0246-1}
  {\bibfield  {journal} {\bibinfo  {journal} {Nat. Phys.}\ }\textbf {\bibinfo
  {volume} {14}},\ \bibinfo {pages} {925} (\bibinfo {year} {2018})}\BibitemShut
  {NoStop}%
\bibitem [{\citenamefont {Cerjan}\ \emph {et~al.}(2020)\citenamefont {Cerjan},
  \citenamefont {J\"urgensen}, \citenamefont {Benalcazar}, \citenamefont
  {Mukherjee},\ and\ \citenamefont {Rechtsman}}]{Cerjan21prl}%
  \BibitemOpen
  \bibfield  {author} {\bibinfo {author} {\bibfnamefont {A.}~\bibnamefont
  {Cerjan}}, \bibinfo {author} {\bibfnamefont {M.}~\bibnamefont {J\"urgensen}},
  \bibinfo {author} {\bibfnamefont {W.~A.}\ \bibnamefont {Benalcazar}},
  \bibinfo {author} {\bibfnamefont {S.}~\bibnamefont {Mukherjee}}, \ and\
  \bibinfo {author} {\bibfnamefont {M.~C.}\ \bibnamefont {Rechtsman}},\ }\href
  {\doibase 10.1103/PhysRevLett.125.213901} {\bibfield  {journal} {\bibinfo
  {journal} {Phys. Rev. Lett.}\ }\textbf {\bibinfo {volume} {125}},\ \bibinfo
  {pages} {213901} (\bibinfo {year} {2020})}\BibitemShut {NoStop}%
\bibitem [{\citenamefont {Christian}\ and\ \citenamefont
  {Immanuel}(2017)}]{Christian17Science}%
  \BibitemOpen
  \bibfield  {author} {\bibinfo {author} {\bibfnamefont {G.}~\bibnamefont
  {Christian}}\ and\ \bibinfo {author} {\bibfnamefont {B.}~\bibnamefont
  {Immanuel}},\ }\href {\doibase 10.1126/science.aal3837} {\bibfield  {journal}
  {\bibinfo  {journal} {Science}\ }\textbf {\bibinfo {volume} {357}},\ \bibinfo
  {pages} {995} (\bibinfo {year} {2017})}\BibitemShut {NoStop}%
\bibitem [{\citenamefont {Tarruell}\ \emph {et~al.}(2012)\citenamefont
  {Tarruell}, \citenamefont {Greif}, \citenamefont {Uehlinger}, \citenamefont
  {Jotzu},\ and\ \citenamefont {Esslinger}}]{Tarruell12nature}%
  \BibitemOpen
  \bibfield  {author} {\bibinfo {author} {\bibfnamefont {L.}~\bibnamefont
  {Tarruell}}, \bibinfo {author} {\bibfnamefont {D.}~\bibnamefont {Greif}},
  \bibinfo {author} {\bibfnamefont {T.}~\bibnamefont {Uehlinger}}, \bibinfo
  {author} {\bibfnamefont {G.}~\bibnamefont {Jotzu}}, \ and\ \bibinfo {author}
  {\bibfnamefont {T.}~\bibnamefont {Esslinger}},\ }\href {\doibase
  10.1038/nature10871} {\bibfield  {journal} {\bibinfo  {journal} {Nature}\
  }\textbf {\bibinfo {volume} {483}},\ \bibinfo {pages} {302} (\bibinfo {year}
  {2012})}\BibitemShut {NoStop}%
\bibitem [{\citenamefont {Shao}\ \emph {et~al.}(2021)\citenamefont {Shao},
  \citenamefont {Liu}, \citenamefont {Xiao}, \citenamefont {Yang},\ and\
  \citenamefont {Zhao}}]{Shao21prl}%
  \BibitemOpen
  \bibfield  {author} {\bibinfo {author} {\bibfnamefont {L.~B.}\ \bibnamefont
  {Shao}}, \bibinfo {author} {\bibfnamefont {Q.}~\bibnamefont {Liu}}, \bibinfo
  {author} {\bibfnamefont {R.}~\bibnamefont {Xiao}}, \bibinfo {author}
  {\bibfnamefont {S.~A.}\ \bibnamefont {Yang}}, \ and\ \bibinfo {author}
  {\bibfnamefont {Y.~X.}\ \bibnamefont {Zhao}},\ }\href {\doibase
  10.1103/PhysRevLett.127.076401} {\bibfield  {journal} {\bibinfo  {journal}
  {Phys. Rev. Lett.}\ }\textbf {\bibinfo {volume} {127}},\ \bibinfo {pages}
  {076401} (\bibinfo {year} {2021})}\BibitemShut {NoStop}%
\bibitem [{\citenamefont {Xue}\ \emph {et~al.}(2021)\citenamefont {Xue},
  \citenamefont {Wang}, \citenamefont {Huang}, \citenamefont {Cheng},
  \citenamefont {Yu}, \citenamefont {Foo}, \citenamefont {Zhao}, \citenamefont
  {Yang},\ and\ \citenamefont {Zhang}}]{Xue21arXiv}%
  \BibitemOpen
  \bibfield  {author} {\bibinfo {author} {\bibfnamefont {H.}~\bibnamefont
  {Xue}}, \bibinfo {author} {\bibfnamefont {Z.}~\bibnamefont {Wang}}, \bibinfo
  {author} {\bibfnamefont {Y.-X.}\ \bibnamefont {Huang}}, \bibinfo {author}
  {\bibfnamefont {Z.}~\bibnamefont {Cheng}}, \bibinfo {author} {\bibfnamefont
  {L.}~\bibnamefont {Yu}}, \bibinfo {author} {\bibfnamefont {Y.~X.}\
  \bibnamefont {Foo}}, \bibinfo {author} {\bibfnamefont {Y.~X.}\ \bibnamefont
  {Zhao}}, \bibinfo {author} {\bibfnamefont {S.~A.}\ \bibnamefont {Yang}}, \
  and\ \bibinfo {author} {\bibfnamefont {B.}~\bibnamefont {Zhang}},\
  }\href@noop {} {} (\bibinfo {year} {2021}),\ \Eprint
  {http://arxiv.org/abs/2107.11564} {arXiv:2107.11564 [cond-mat.mes-hall]}
  \BibitemShut {NoStop}%
\bibitem [{\citenamefont {Tam}\ \emph {et~al.}(2022)\citenamefont {Tam},
  \citenamefont {Venderbos},\ and\ \citenamefont {Kane}}]{Tam22prb}%
  \BibitemOpen
  \bibfield  {author} {\bibinfo {author} {\bibfnamefont {P.~M.}\ \bibnamefont
  {Tam}}, \bibinfo {author} {\bibfnamefont {J.~W.~F.}\ \bibnamefont
  {Venderbos}}, \ and\ \bibinfo {author} {\bibfnamefont {C.~L.}\ \bibnamefont
  {Kane}},\ }\href {\doibase 10.1103/PhysRevB.105.045106} {\bibfield  {journal}
  {\bibinfo  {journal} {Phys. Rev. B}\ }\textbf {\bibinfo {volume} {105}},\
  \bibinfo {pages} {045106} (\bibinfo {year} {2022})}\BibitemShut {NoStop}%
\end{thebibliography}%
\end{document}